\documentstyle[prb,aps,epsfig,multicol]{revtex}
\begin{document}
\newcommand{\bra}[1]{\langle #1|}
\newcommand{\ket}[1]{|#1\rangle}
\newcommand{\braket}[1]{\langle#1|#1\rangle}
\sloppy
\draft
\title
{Jahn-Teller polarons and their superconductivity in a molecular
conductor}
\author{R. Ramakumar and Sudhakar Yarlagadda}
\address{Condensed Matter Physics Group, Saha Institute of Nuclear Physics,
1/AF Bidhannagar, Calcutta-700064, INDIA} 
\date{29 July 2003}
\maketitle
\begin{abstract}
We present a theoretical study of a possibility of superconductivity 
in a three dimensional molecular conductor in which the interaction between 
electrons in doubly degenerate molecular orbitals
and an {\em intra}molecular vibration mode is large 
enough to lead to the formation of $E\otimes \beta$ Jahn-Teller small polarons.
We argue that the effective polaron-polaron interaction can be
attractive for material parameters realizable in molecular conductors.
This interaction is the source of superconductivity in our model.
On analyzing superconducting instability in the weak and strong
coupling regimes of this attractive interaction, 
we find that superconducting transition temperatures up to 100 K 
are achievable in molecular conductors within this mechanism.
We also find, for two particles per molecular site,  a novel Mott 
insulating state in which a polaron singlet occupies one of
the doubly degenerate orbitals on each site. Relevance of 
this study in the
search for new molecular superconductors is pointed out.
\end{abstract}


\begin{multicols}{2}
\narrowtext
\section{Introduction}
\label{sec1}

Ever since the experimental discovery\cite{jerome1} of 
superconductivity by Jerome {\em et al.,}
in 1980 in a Bechgaard\cite{bech} 
salt (TMTTF)$_2$PF$_6$ (with a T$_c$ of 1.2 K under a pressure of 6.5 Kbar),
search for new higher T$_c$ molecular superconductors has been a vigorous
field of research\cite{yam1}. 
A large number of molecular superconductors\cite{yam1}
with varying degrees of dimensionality
have been discovered since 1980, and
a continuing multidisciplinary search for new higher T$_c$ molecular
superconductors is currently underway. 
Theoretical studies of a possibility of superconductivity in 
molecular conductors and conducting polymers have had a positive influence
in the development of this field.  
Indeed, Little's\cite{little}
prediction of high superconducting transition temperatures within an
exciton mediated mechanism of
superconductivity
in a hypothetical conducting polymer chain (with polarizable molecules
periodically attached to the spine) played a boosting 
role for this field\cite{jerome2}.
For a discussion on the future 
prospects for superconductivity in conducting polymers see  Heeger  
\cite{heeger}.
\par
One of the strategies in the search for higher T$_c$ molecular
superconductors continues to be to
search for molecules with large electron-{\em intra}molecular vibration
coupling (EMVC) and to crystallize them with the hope that the solid will
be metallic or can be made metallic by charge transfer from another molecule
(in charge-transfer salts) and/or by application of pressure. If the
material is metallic and if EMVC is the pairing glue, large EMVC implies
a large T$_c$ within the frame-work of BCS theory\cite{bcs}. In a molecular
conductor, the phonon spectrum forms two distinct groups: high-frequency
{\em intra}molecular phonons and {\em inter}molecular phonons which have
relatively low frequencies ($< 300$ cm$^{-1}$ in the currently known organic
superconductors). Where calculations are available\cite{graja}, the EMVC's are larger
than the electron-{\em inter}molecular vibration coupling, though there
are exceptions to this trend\cite{ram}.
But, when EMVC's become large, there is a
possibility of polaron formation\cite{alex}, a
possibility not included in the
Migdal-Eliashberg\cite{mig,eli} extension of the BCS theory.
If EMVC is large enough to lead
to small polaron formation, a possibility for superconductivity to arise
is by the polaron pairing through the exchange
of low frequency {\em inter}molecular phonons.
Recently we investigated\cite{ram1} this possibility
in a simple model molecular 
conductor. In that work we considered molecular orbitals, overlaps 
of which on nearest neighbor (NN) molecules produces the conduction band,
to have {\em no orbital degeneracy}.
\par
Now, depending on the symmetry of the 
molecule, the molecular orbitals which participate in the band formation
can be degenerate or non-degenerate. The purpose of the present paper
is to investigate the possibility of superconductivity in a three dimensional
molecular conductor in which the molecular orbitals participating
in the conduction band formation 
are doubly degenerate and EMVC between electrons and
a single 
{\em intra}molecular vibration mode is large
enough to lead to the formation of Jahn-Teller (JT) small polarons.
This is the $E\otimes \beta$ JT model and is the simplest
JT system. Molecules with tetragonal symmetry are examples 
of such a system.
\par
The rest of the paper is organized as follows. In Sec. II, we introduce
our model for the molecular conductor, perform a Lang-Firsov transformation
on it to obtain the JT polaron Hamiltonian. 
In Sec. III, a BCS-mean-field (BCS-MF) theory is presented which is
applicable in the weak coupling regime of the effective attractive interaction
between the JT small polarons. In Sec. IV, we analyze the strong coupling Bose
regime. An insulating state obtained for the
case of two electrons per site is discussed in Sec. V. The conclusions
are given in Sec. VI.
\section{Jahn-Teller small polarons and their effective interactions}
\label{sect2}
We employ the following model to study the $E\otimes \beta$
JT polaron formation and
their effective interactions:
\begin{eqnarray}
\displaystyle
H&=&t\sum_{ij}\sum_{\alpha\sigma}
c^{\dag}_{i\alpha\sigma}c_{j\alpha\sigma}
+\omega\sum_{\bf{q}}b^{\dag}_{\bf{q}}b_{\bf{q}}
\nonumber \\
&+&\frac{G}{\sqrt{N}}\sum_{j\bf{q}\sigma}(n_{j1\sigma}-n_{j2\sigma})
e^{i\bf{q}.\bf{R_{i}}}(b^{\dag}_{\bf{-q}}+b_{\bf{q}})
\nonumber \\
&+&U\sum_{j\alpha}n_{j\alpha\uparrow}n_{j\alpha\downarrow}
+U\sum_{j\sigma\sigma^{\prime}}
n_{j1\sigma}n_{j2\sigma^{\prime}}\,.
\end{eqnarray}
In the above $H$, $i$ and $j$ are the molecular site indices, $\alpha$
(=1, 2) is the orbital index of the two degenerate molecular orbitals,
$t$ ($<0$) the hopping energy between similar
orbitals on NN molecules, $\omega$ the molecular vibration frequency, $G$
the electron-phonon ({\em el-ph}) interaction energy,
$M$ the mass, $U$ the
Coulomb interaction strength, and $N$ the number of sites.
Furthermore, $c_{j\alpha\sigma}$ and $b_{\bf q}$ are the
destruction operators of electrons and phonons respectively, 
and $n_{j\alpha\sigma}=c^{\dag}_{j\alpha\sigma}c_{j\alpha\sigma}$.
A chemical potential will be introduced later when we have to fix the
electron density. In the Appendix, using an argument
given in Ref.\cite{sturge} we show that, for $2G/\omega\gg 1$,
the  $E\otimes e$ JT model
reduces to $E\otimes \beta$ JT model. A detailed numerical
study of polaron and bipolaron formation in $E\otimes e$ JT
model in one dimension and for the case of one and two electrons
was recently published by Shawish {\em et.al.}\cite{trugman}.
\par
Our calculation proceeds along the following lines. 
First we apply a multi-band Lang-Firsov (LF) transformation\cite{lf} which produces 
the JT small polarons, reduces
the {\em intra}orbital Coulomb repulsion, and increases the {\em inter}orbital
Coulomb repulsion. We will argue that the {\em intra}orbital polaron-polaron
interaction can be attractive for realistic range 
of frequencies and {\em el-ph} coupling realizable in
molecular conductors. This interaction then is the
source of superconductivity in our model. In the next step, we project out
{\em inter}orbital double occupancies employing a
Gutzwiller approximation\cite{gutz}
method in the weak and strong coupling regimes of the attractive interaction
to obtain effective Hamiltonians in these regimes. 
These effective JT Hamiltonians are used to
study superconductivity in the weak-coupling BCS-MF and strong-coupling (Bose)
regimes. The superconducting transition temperatures
obtained are then 
analyzed as a function of phonon frequency,
{\em el-ph} interaction strength, Coulomb repulsion strength, original
band-width, and doping. We will also see that, for the case of two polarons per site,
we obtain a novel insulating state with a polaron singlet occupying the lattice
sites. For this paired polaron Mott insulator, 
we show that one obtains a novel Orbital Density Wave State (ODW)  for 
large and finite {\em inter}orbital repulsion.
We also point out a mapping of this ODW sate to the 
Ising model. From our analysis of the superconducting transition temperature
in the BCS-MF regime and the Bose regime, it will be shown that  
superconducting T$_c$ up to 100 K is achievable for this mechanism.
Now we go to the details of the calculations. 
\par
As stated, we apply a multi-band LF
transformation to the $H$. The transformation $e^{S}He^{-S}=
H_T$ transforms $H$ into the polaron representation.
LF is a unitary transformation with $S=S_{1}+S_{2}$, where $S_{\alpha}$
($\alpha=1,\,2$) is 
\begin{equation}
S_{\alpha}=(-1)^{\alpha}\frac{G}{\omega}\frac{1}{\sqrt{N}}
\sum_{j\bf{q}\sigma}n_{j\alpha\sigma}
e^{i{\bf q}.{\bf R}_{j}}(b_{{\bf q}}-b^{\dag}_{-{\bf q}}).
\end{equation}
Using this $S$, we have
\begin{equation}
\displaystyle
e^{S}b^{\dag}_{\bf{q}}e^{-S}=b^{\dag}_{\bf{q}}-
\frac{G}{\omega}\frac{1}{\sqrt{N}}
\sum_{j\sigma}(n_{j1\sigma}-n_{j2\sigma})e^{i{\bf q}.{\bf R}_{j}}\,,
\end{equation}
\begin{equation}
e^{S}c_{j\alpha\sigma}e^{-S}=e^{S_{\alpha}}c_{j\alpha\sigma}e^{-S_{\alpha}
}=c_{j\alpha\sigma}X_{j\alpha}\,,
\end{equation} 
and 
\begin{equation}
X_{j\alpha}=exp\left[(-1)^{\alpha-1}\frac{G}{\omega}
\frac{1}{\sqrt{N}}
\sum_{\bf{q}}e^{i\bf{q.R_{j}}}(b_{{\bf q}}-b^{\dag}_{-{\bf q}}) \right]\,.
\end{equation}
Using the above operators, the transformed Hamiltonian is
\begin{equation}
H_{T}=H_{KE}+H_{R}+H_{A}+H_{I}\,,
\end{equation}
where
\begin{equation}
H_{KE}=t\sum_{ij}\sum_{\alpha\sigma}
c^{\dag}_{i\alpha\sigma}c_{j\alpha\sigma}
X^{\dag}_{i\alpha}X_{j\alpha}\,,
\end{equation}
\begin{equation}
H_{R}=U_{R}\sum_{j\sigma\sigma^{\prime}}n_{j1\sigma}n_{j2\sigma^{\prime}}\,,
\end{equation}
\begin{equation}
H_{A}=U_{A}\sum_{j\alpha}n_{j\alpha\uparrow}n_{j\alpha\downarrow} \,,
\end{equation}
and
\begin{equation}
H_{I}=\omega\sum_{\bf{q}}b^{\dag}_{\bf{q}}b_{\bf{q}}
-\frac{G^{2}}{\omega}
\sum_{j\alpha\sigma}n_{j\alpha\sigma}\,.
\end{equation}
In the above equations
\begin{equation}
U_{R}=U+2\frac{G^{2}}{\omega}\,, 
\end{equation}
and
\begin{equation}
U_{A}=U-2\frac{G^{2}}{\omega}\,.
\end{equation}
The LF transformation produces three effects. The first is the polaronic
band-width reduction, of course. The second and third effects are the
enhancement of the {\em inter}orbital Coulomb repulsion and the reduction
of the {\em intra}orbital Coulomb repulsion, respectively. These
second and third effects have different origins: while the second
has its origin in the JT nature of the system, the third
is a small polaron effect. On phonon
averaging\cite{sy} of $H_{T}$ as usual, neglecting a constant
coming from $H_I$, one  obtains
\begin{equation}
H_{P}=H^{P}_{KE}+H_{R}+H_{A}\,,
\end{equation}
where
\begin{equation}
H^{P}_{KE}=t_{P}\sum_{ij}\sum_{\alpha\sigma}
c^{\dag}_{i\alpha\sigma}c_{j\alpha\sigma}\,,
\end{equation}
in which
\begin{equation}
t_{P}=t\times 
exp\left[-\left(\frac{G}{\omega}\right)^{2}
coth\left(\frac{\beta\omega}{2}\right) \right]\,,
\end{equation}
and $\beta=1/k_{B}T$. 
\par
Now, $H_{P}$ is the Hamiltonian for a collection of JT small polarons and these
polarons interact among themselves though in the {\em intra} and {\em inter}
orbital interactions $U_{A}$ and $U_R$, respectively. At this stage, we 
demand that $U_{A}$ be an attractive interaction. This is very much
realizable for the values of $U$, $G$, and $\omega$ in a
range possible in molecular conductors. Some idea about the range
of $U$, $G$, and $\omega$ would be useful to get a better feel.
The highest {\em intra}molecular vibration frequency is unlikely to exceed the
frequency ($4161$ cm$^{-1}$) of the lightest molecule (H$_2$, the Hydrogen molecule).
Among the calculated values\cite{yamabe}
of $G$, the highest seems to be around
$200$ meV (obtained for a mode of frequency $1656$ cm$^{-1}$ in
Benzene). Practically all the existing molecular superconductors
are built from the
molecules TMTSF, BEDT-TTF, DMIT, DMET, BETS, or C$_{60}$. This is 
arguably a small subset of already synthesizable molecules.
Then, it is quite possible that higher values of $G$ are realized
in many molecules. As for $U$, large and  highly polarizable molecules can
be expected to have small $U$'s. In the existing molecular superconductors
\cite{yam1}, $U$ seems to be in a range around $1$ eV. 
We note that the $U$ is {\em not} for an isolated molecule, but
for a molecule sitting on a lattice site in the solid so that
$U$ includes the effects of polarization of molecules surrounding
a given molecule and consequently is reduced from the $U$ value
for the isolated molecule.
Now, if we tentatively fix an upper limit for $G$ around $600$ meV and keep
the upper limit of $\omega$ as $4161$ cm$^{-1}$,  
it is not unreasonable to expect $U_{A}$ to be negative in a range of $U$
which obtains in molecular conductors. Given the estimates of {\em el-ph}
coupling to individual modes, this situation is very unlikely to be realized
in the existing organic superconductors. But, that does not preclude its
realizability in another important class of materials: the molecular
conductors yet to be discovered. In our study, we assume that $U_{A}$ is
negative. This attractive $U_{A}$ is then the origin of superconductivity
in our theory. In the next two sections, we study superconductivity in the
weak and strong coupling regimes of $U_A$. 
\section{BCS mean-field theory in the weak coupling regime of $U_A$}
\label{sect3}
The weak coupling regime of $U_A$ is the range of $U_A$ in which the
superconducting T$_c$ and the pair formation temperature are the same.
In this section, we project out
{\em inter}orbital double occupancies, using the Gutzwiller approximation
method, to obtain an effective Hamiltonian in the weak coupling regime
of $U_A$ and then study the superconductivity using a BCS-MF theory.
The Gutzwiller projection renormalizes $t_P$ and $U_A$, and decouples
the orbitals. 
First, we recall that the bare {\em el-el} repulsions [in Eq. (1)] are
strongly renormalized [Eqs. (11) and (12)]: while the {\em inter}orbital 
repulsions have increased, the {\em intra}orbital repulsion is suppressed to
an extent that it has become attractive (see the previous paragraph).
The effect of $U_R$ is to strongly reduce the probability of polarons 
occupying different orbitals on a given site. On the other hand, $U_A$
(now attractive) increases the probability of two polarons (with opposite
spins) occupying the same orbital on a given site. Both these aspects have
to be considered while doing the Gutzwiller projection. We take, for
simplicity, the $U_R$ to be infinity and completely project out the 
{\em inter}orbital double occupancies. Qualitatively one can see
that projecting the {\em inter}orbital double occupancies reduces the
polaron band-width (through a density dependent function) since 
it is energetically unfavorable for a polaron to hop from an orbital
(say 1) on a site to a target site where the orbital 2 is occupied.
Also, strong reduction of 
{\em inter}orbital double occupancies increases the {\em intra}orbital 
double occupancies and consequently modifies the {\em intra}orbital
attraction. We do the {\em inter}orbital double occupancy projection
in an approximate way. First consider the part $H^{P}_{KE}+H_R$ of
$H_P$ [Eq. (13)]. We write the Gutzwiller wave function 
$\ket{\psi_{GW}}$ as:
\begin{equation}
\ket{\psi_{GW}}=\prod_{i\sigma\sigma^{\prime}}
[1-(1-\eta)n_{i1\sigma}n_{i2\sigma^{\prime}}]\ket{\psi_{\circ}} \,,
\end{equation}
where
\begin{equation}
\ket{\psi_{\circ}}=\prod_{k_{1}<k_{F}}c^{\dag}_{k1}\ket{0}\otimes
                    \prod_{k_{2}<k_{F}}c^{\dag}_{k2}\ket{0}\,.
\end{equation}
\par
For calculating the expectation value of $H^{P}_{KE}$ and $H_R$ in
$\ket{\psi_{GW}}$, we use a simple intuitive method 
proposed by Okabe\cite{okabe}.
Since the maximum number of polarons 
($\sum_{\alpha\sigma}n_{\alpha\sigma}$)
a site can accommodate is four, there are sixteen possibilities for
occupancy for finite $U_R$. As mentioned earlier, we take $U_R$ infinity
limit for simplicity [$\eta=0$ in Eq. (16)],
and consequently the sixteen possibilities reduces
to seven. In the paramagnetic case
($i.\,e.,\,n_{1\uparrow}=n_{1\downarrow}=n_{2\uparrow}=
n_{2\downarrow}=n/4)$,
the probability for a site to be vacant is $[1-(n-2d)]$ and the 
probability for a site to be singly occupied is $(n/4-d)$. Here $d$
is the {\em intra}orbital double occupancy. Then, considering the hopping
process given in Fig. 1, the Gutzwiller band narrowing factor is
\begin{equation}
q(n,d)=\frac{\left[\sqrt{(\frac{n}{4}-d)(1-n+2d)}
+\sqrt{d(\frac{n}{4}-d)}\,\right]^2}{\frac{n}{4}(1-\frac{n}{4})}\,,
\end{equation}
where the numerator and the denominator corresponds to hopping in the presence
and absence of 
{\em inter}orbital repulsion, respectively.
The ground state energy for $H^{P}_{KE}+H_R$ is
\begin{equation}
E_{c}(n,d)=q(n,d)\sum_{\alpha\sigma}\epsilon_{\alpha\sigma} \,,
\end{equation}
where $\epsilon_{\alpha\sigma}[=\epsilon]$ is the average kinetic energy
per polaron, and is independent of the indices $\alpha$ and $\sigma$.
Minimization of $E_{c}(n,d)$ with respect to $d$ leads to
the equation to determine $d$ for a given $n$:
\begin{equation}
8d^{3}+(4-6n)d^{2}+(1-2n+\frac{9}{8}n^{2})d+\frac{n^{3}-n^{2}}{16}=0\,.
\end{equation}
For $n=2$, the above equation implies that $d=0.5$, which means that
one of the orbitals on all the sites are occupied by
a singlet. For a general value of $n$, one has to solve Eq. (20) to obtain
the variation of $d$ as a function of $n$. The optimized values of
the band narrowing factor
$q(n,d_{opt})$ and the {\em intra}orbital double occupancy ($d_{opt}$)
as a function of $n$ are shown in Fig. 2 and Fig. 3, respectively.
From the numerical results shown in the figures, we have
\begin{equation}
q(n)=q(n,d_{opt})\approx\frac{2-n}{2}\,, \\
\end{equation}
and
\begin{equation}
d_{opt}\approx\frac{n^2}{8}\,.
\end{equation}
Furthermore, it is clear that
as $n\rightarrow 2$, the polaron band-widths are narrowed and a Mott
insulator is obtained for $n=2$. This Mott insulator has a singlet
in one of the orbitals on 
each lattice site. The properties and excitations of this insulator
will be discussed separately  in Sect. V. Next consider the term $H_A$
in Eq. (13). To get the total ground state energy, we add to $E_{c}(n,d)$
the contribution from $H_A$ obtained using $\ket{\psi_{GW}}$ calculated
for optimized $d$ ($d_{opt}$). This contribution per site is
\begin{equation}
\displaystyle
-|U_{A}|
\frac{\bra{\psi_{GW}}\sum_{\alpha}n_{i\uparrow\alpha}n_{i\downarrow\alpha}
\ket{\psi_{GW}}}
{\braket{\psi_{GW}}} \\
=-2|U_A|d_{opt}\,.
\end{equation}
So, the total minimized energy per site is
\begin{equation}
E(n)=\sum_{\alpha\sigma}q(n,d_{opt})\epsilon_{\alpha\sigma}
-2|U_A|d_{opt}\,.
\end{equation}
Using Eqs. (21) and (22), the minimized energy per site is 
\begin{equation}
E(n)\approx 2(2-n)\epsilon-\frac{|U_A|n^{2}}{4}.
\end{equation}
%
\par 
Using the results of the Gutzwiller projection of inter-orbital
double occupancies, we write an effective Hamiltonian for the JT
polarons as:
\begin{equation}
\tilde{H}_{P}=q(n)t_{P}\sum_{ij}\sum_{\alpha\sigma}
c^{\dag}_{i\alpha\sigma}c_{j\alpha\sigma}+
\tilde{U}_{A}\sum_{j\alpha}n_{j\alpha\uparrow}n_{j\alpha\downarrow}\,.
\end{equation}
The effects of Gutzwiller projection are the density dependent
band-width renormalization and a renormalization of the on-site 
{\em intra}orbital attractive interaction from $U_A$ to $\tilde{U}_A$.
From Eq. (25), it follows that $\tilde{U}_{A}=2U_A$. We notice
that this enhancement is similar to the enhancement of 
antiferromagnetic exchange interaction ($J$) obtained on Gutzwiller
projection in a $t-U-J$ model studied in Ref\cite{zhang}.
For the $t-U-J$ model, they found that the bare $J$ steadily
increases with increasing electron density to reach $4J$
in the Mott insulator at the half-filling of their non-degenerate band.
Notice also, for our model,
that orbitals are decoupled on Gutzwiller projection.
\par
Now, $\tilde{H}$ is clearly an effective attractive Hubbard model. We
note that, unlike the usual phonon mediated attractive interaction
which is attractive only in the Debye shell around the Fermi surface, all
the polarons experience the attraction $\tilde{U}_A$ or $U_A$.
In the next section, we study superconductivity in the weak 
coupling regime of $\tilde{U}_A$ using
a BCS type mean-field (BCS-MF) analysis to obtain superconducting
energy gap and $T_c$.
\par
To move forward, we have to fix the weak and strong coupling regimes
of the single band attractive Hubbard model
($H_{AHM}=t\sum_{ij\sigma}
c^{\dag}_{i\sigma}c_{j\sigma}-
|U|\sum_{j}n_{j\uparrow}n_{j\downarrow}$)
in terms of $|U|/t$.
We can approximately fix these ranges using the existing  
studies\cite{mic,met,beck} on three dimensional non-degenerate 
attractive Hubbard model. 
A comparison of results for superconducting $T_c$ 
obtained using three methods (T-Matrix Approximation, Determinant
Quantum Monte Carlo, and Dynamical Mean Field Theory) and 
the BCS-MF is given in Fig. 2 of Ref.\cite{beck} 
Notice that these
results are for quarter filling of the band. 
The regime
of weak coupling, for a given $n$, is the range of $|U|/t$ within
which the pair formation temperature and the superconducting
transition temperature are the same. Using the calculation of
Pauli susceptibility (which would be suppressed above the
superconducting $T_c$ in the case pairs form above $T_c$),
the authors of\cite{beck} argued that $0<|U|/t<4$ is the BCS-MF
regime for $n=0.5$. Considering together all these results, we tentatively
fixed the BCS-MF regime of the 3D attractive Hubbard model to be
$0<|U|/t<3$. As for the strong coupling regime, all the electrons
are in paired states below a pair formation temperature of $O(|U|)$.
We fixed the Bose regime to be for $|U|/t>12$. The intermediate 
coupling regime then is obviously $3<|U|/t<12$. 
In the weak coupling regime of $U_A$, the ratio equivalent 
to $|U|/t$ in our case is, from Eq. (26), $|\tilde{U}_{A}|/q(n)t_{P}$.
In the strong coupling regime of $U_A$ studied in the next section,
the Gutwiller projection
has to be done in a different way than that given in this section,
and that gives $|U_A|/\sqrt{\bar{q}(n)}t_P$ [where $\bar{q}(n)=(1-n/2)/
(1-n/4)$] as the ratio equivalent to $|U|/t$.
We can now
proceed to the analysis of superconductivity in the BCS-MF  
regime of the effective attractive interaction $\tilde{U}_A$.
\par
We have from Eq. (26), 
\begin{equation}
\tilde{H}^{\alpha}_{P}=\sum_{{\bf k}\sigma}
\xi^{\alpha}_{P}({\bf k})n_{{\bf k}\alpha\sigma}
+\tilde{U}_{A}\sum_{j}n_{j\alpha\uparrow}n_{j\alpha\downarrow}\,, 
\end{equation}
where
\begin{equation}
\xi^{\alpha}_{P}(k)=q(n)\epsilon^{\alpha}_{P}(k)-\mu\,, 
\end{equation}
and 
\begin{equation}
\epsilon^{\alpha}_{P}(k)=2t_{P}[cos(k_{x}a)+cos(k_{y}a)+cos(k_{z}a)]\,,
\end{equation}
for a simple cubic lattice. We have introduced a chemical potential ($\mu$)
to fix the number density. 
Given the form of $\tilde{U}_{A}$, only isotropic s-wave 
pairing is possible. A BCS-MF theory of $\tilde{H}^{\alpha}_{P}$
gives:
\begin{equation}
\tilde{H}^{\alpha}_{P}=\sum_{{\bf k}\sigma}
\xi^{\alpha}_{P}({\bf k})n_{{\bf k}\alpha\sigma}
+\sum_{{\bf k}}(\Delta^{\star}_{\alpha}
c_{-{\bf k}\alpha\downarrow}c_{{\bf k}\alpha\uparrow}
+h.c.)\,,
\end{equation}
where the superconducting gap parameter is
\begin{equation}
\Delta^{\star}_{\alpha}=\tilde{U}_{A}\sum_{{\bf k}}
\left<c^{\dag}_{{\bf k}\alpha\uparrow}c^{\dag}_{-{\bf k}\alpha\downarrow}\right>\,.
\end{equation}
The superconducting gap and the particle density is determined from
\begin{equation}
\displaystyle
\frac{2}{\left|\tilde{U}_{A}\right|}=\int_{-q(n)D_{P}}^{+q(n)D_{P}}
\frac{N_{P}(\epsilon^{\alpha})}{q(n)}
F(\epsilon^{\alpha},\Delta_{\alpha},\mu,\beta)\,d\epsilon^{\alpha}\,,
\end{equation}
and
\begin{equation}
\displaystyle
1-n_{\alpha}=\int_{-q(n)D_{P}}^{+q(n)D_{P}}
\frac{N_{P}(\epsilon^{\alpha})}{q(n)}(\epsilon^{\alpha}-\mu)
F(\epsilon^{\alpha},\Delta_{\alpha},\mu,\beta)\,d\epsilon^{\alpha}\,,
\end{equation}
where
\begin{equation}
F(\epsilon^{\alpha},\Delta_{\alpha},\mu,\beta)
=\frac{tanh\left(.5\beta\sqrt{(\epsilon_{\alpha}-\mu)^2
+|\Delta_{\alpha}|^2}\right)}
{\sqrt{(\epsilon_{\alpha}-\mu)^2+|\Delta_{\alpha}|^2}}\,.
\end{equation}
Here $N_{P}(\epsilon^{\alpha})$ is the polaron Density Of States (DOS)
corresponding to the polaron band $\epsilon^{\alpha}_{P}(k)$. 
Assuming a square one-polaron single-spin DOS ({\em i.e.}, 
$N_{P}(\epsilon^{\alpha})=1/2D_{P}=1/12t_{P})$, the solutions of
the gap and chemical potential equations give:
\begin{equation}
\Delta_{\alpha}(T=0)=\frac{\sqrt{\frac{\delta^2}{4}-\frac{\delta^4}{16}}
\,D_{P}}{sinh\left(\frac{1}{\lambda}\right)}\,,
\end{equation}
and
\begin{equation}
k_{B}T_{c}=\sqrt{\frac{\delta^2}{4}-\frac{\delta^4}{16}}
\,1.13D_{P}e^{-\frac{1}{\lambda}}\,.
\end{equation}
In the above equations, $\lambda=2|U-2g^{2}\omega|/(2q(n)D_{P})$, 
$D_{P}=D_{\circ}e^{-g^{2}}$, $\delta=2-n$,
$g=G/\omega$, and $n$ is the number of polarons per site.
For the values of $g$ and $\omega$ we use in numerical calculations,
the temperature dependence of the polaron band-width is 
practically nil for the the temperature range ($<300$ K) of interest
to us.
The equation for $T_c$ is valid for $[(q(n)D_{P}\pm\mu)/k_{B}T_c]\gg 1$.
See also the note in Ref\cite{rice}.
\par
Now we go to the estimates of $\Delta_{\alpha}(0)$ and $T_C$ considering
realistic values of $g,\,\omega,\,D_{\circ}$, 
and $U$ applicable to molecular
conductors. While making the estimates, one has to satisfy several
constraints. These are: (i) JT small polaron formation condition 
($g^{2}\omega>q(n)D_{\circ}$); (ii) Phonon  frequency should 
be below the frequency ($4161$ cm$^{-1}$) of the lightest molecule (H$_2$);
(iii) A realistic upper limit for {\em el-ph} interaction strength was
earlier fixed at $600$ meV; and (iv) Range of validity of BCS-MF theory 
requires that $\lambda<0.25$. These conditions enforce severe constraints
on the parameters $g,\,\omega,\,D_{\circ}$, and $U$ for which
the system can undergo superconducting instability and if that
happens on the achievable $T_c$'s. 
The variation of $\Delta_{\alpha}$ (at $T=0$) with $\delta$
is shown in Figs. 4-6,
for different values of $\omega$, $D_{o}$, and $g$, respectively.
Since the trends are similar for $\Delta_{\alpha}$ and T$_c$, we will discuss
the T$_c$ curves shown in Figs. 7-9.
The lower limit of $\delta$ for the curves in Figs. 4-9 is determined
by the condition on $\lambda$, and the upper cut-off as a function
of $\delta$ enforces the small polaron condition.
The variation of $T_c$ with $\delta$
for various values of $\omega$ is shown in Fig. 7. For fixed $U$, 
$D_\circ$ and $g$, BCS-MF superconductivity is confined to the $\omega$
range: $\omega_{min}<\omega<\omega_{max}$
where $\omega_{min}=U/2g^2$ and $\omega_{max}=
\omega_{min}+(0.25q(n)D_{P}/2g^2)$. In this allowed range 
of $\omega$, moderate values of $T_c$ are possible. 
Increasing $\omega$ is found to increase the $T_c$'s achievable and 
expands the density range in which superconductivity is possible.
The variation of $T_c$ with 
$\delta$ for different values of $D_\circ$ is shown in Fig. 8, and
it shows that in the range shown changes in $D_\circ$ do not
have much effect. On the other hand, the $T_c$ vs. $\delta$ curves
shown in Fig. 9 show that small changes in $g$ severely narrow the
range of $\delta$ for which superconductivity is possible. 
It is clear that the various conditions mentioned earlier severely
limit the range of $\delta$ for which superconductivity is possible
and the values of $T_c$'s achievable. We would also like to 
note that below the lower cut-off of $\delta$ for each curve in
the Figs. 4-9, the $\lambda$ moves out  of the BCS-MF regime into
the intermediate to strong coupling regime and
whether superconductivity is possible in that range of $\delta$ 
cannot be addressed using the weak coupling formulas used in
this section. The values of $U$, $g$, $D_\circ$, and $\omega$ 
used in making the $T_c$ estimates are definitely not unrealistic
considering the variety of molecules synthesizable and crystallizable.
Though band-filling control has not yet been achieved in molecular
conductors, such a possibility cannot be ruled out in the future.
Even if band-filling control is not possible, our results show that
JT polarons can undergo superconducting instability with moderate $T_c$'s
in a realistic range of $U$, $g$, $D_\circ$, and $\omega$
realizable in molecular conductors. 
This completes the analysis of superconductivity in the BCS-MF regime
of our model. In the next section, we consider the strong coupling
regime (the Bose regime) of $U_A$.
\section{Bose condensation in the strong coupling regime of $U_A$}
\label{sect4}
In this regime, all the JT polarons are in paired states below 
a temperature of $O(|U_A|)$. To obtain the Bose condensation temperature
of these pre-formed pairs, we start from the Hamiltonian $H_P$ [see 
Eq. (13)]. Since
the largest terms in $H_P$ are $H_R$ and $H_A$ (in the strong coupling
regime), we treat ($H_{R}+H_A$) as the part which determines the ground
state energy and take $H^{P}_{KE}$ as a perturbation. Now, in the 
ground state of ($H_{R}+H_A$), a site has either a singlet occupying
one of the orbitals or is empty. The {\em inter}orbital pairs are
projected out because of the strong {\em inter}orbital repulsion ($U_R$).
When we switch on the hopping term ($H^{P}_{KE}$), the pairs become mobile
through a second order (in $H^{P}_{KE}$ ) process. For the ground state of
($H_{R}+H_A$) we have
\begin{equation}
(H_{R}+H_A)\ket{p}=E_{\circ}\ket{p}\,,
\end{equation}  
where $E_{\circ}=(-|U_A|/2)\sum_{j\alpha\sigma}n_{j\alpha\sigma}$ is the ground
state energy. The second order pair hopping term then is,
\begin{eqnarray}
E_{2}&=&-\frac{t^{2}_{P}}{|U_A|}\sum_{r}\bra{p}
\sum_{ij}\sum_{\alpha\sigma}c^{\dag}_{i\alpha\sigma}c_{j\alpha\sigma} 
\ket{r}\bra{r} \nonumber \\
&\times& \sum_{lm}\sum_{\beta\sigma^{\prime}}c^{\dag}_{l\beta\sigma^{\prime}
}c_{m\beta\sigma^{\prime}}\ket{p}\,.
\end{eqnarray}  
In Eq. (38), the $r$ sum is over states other than the degenerate ground
states. Since $\sum_{r}\ket{r}\bra{r}=I-\sum_{p}\ket{p}\bra{p}$ 
and noting that $\bra{p}T\ket{p^{\prime}}=0$, the second order term
becomes
\begin{equation}
E_{2}=-\frac{t^{2}_{P}}{|U_A|}
\bra{p}\sum_{ij}\sum_{\alpha\sigma}c^{\dag}_{i\alpha\sigma}c_{j\alpha\sigma}
\sum_{lm}\sum_{\beta\sigma^{\prime}}c^{\dag}_{l\beta\sigma^{\prime}}
c_{m\beta\sigma^{\prime}}\ket{p}
\,.
\end{equation}
The $E_2$ is non-zero for (A) $l=j;\,i=m;\,\alpha=\beta;\,
\sigma=\sigma^{\prime}$ or (B) $i=l;\,j=m;\,\alpha=\beta;\,
\sigma=-\sigma^{\prime}$. Then,
\begin{eqnarray}
E_{2}&=&-\frac{t^{2}_{P}}{|U_A|}
(
\bra{p}2\sum_{ij\alpha}
c^{\dag}_{i\alpha\uparrow}c^{\dag}_{i\alpha\downarrow}
c_{j\alpha\downarrow}c_{j\alpha\uparrow}\ket{p} \nonumber \\
&+&\bra{p}\sum_{ij\alpha}
c^{\dag}_{i\alpha\sigma}c_{j\alpha\sigma}
c^{\dag}_{j\alpha\sigma}c_{i\alpha\sigma}\ket{p}
)
\,. 
\end{eqnarray}
The hopping processes corresponding to the first and second terms
in $E_2$ are shown in Fig. 10. $E_2$ can be calculated considering
the probability amplitudes for the processes shown in the figure
and we obtained it to be the ground state energy of the Hamiltonian
\begin{eqnarray}
H_{B}&=&-2\frac{t^{2}_{P}\bar{q}(n)}{|U_A|}
\sum_{ij\alpha}c^{\dag}_{i\alpha\uparrow}c^{\dag}_{i\alpha\downarrow}
c_{j\alpha\downarrow}c_{j\alpha\uparrow} \nonumber \\
&+&\frac{t^{2}_{P}\bar{q}(n)}{|U_A|}
\sum_{ij}\sum_{\alpha\sigma}n_{i\alpha\sigma}n_{j\alpha\sigma}
-\frac{Zt^{2}_{P}\bar{q}(n)}{|U_A|}\sum_{j\alpha\sigma}n_{j\alpha\sigma}\,,
\end{eqnarray}
where $\bar{q}(n)=(1-n/2)/(1-n/4)$ is the Gutzwiller projection factor
and $Z$ is the co-ordination number
of a site in the lattice. The Bose condensation temperature ($T_B$)
in the strong coupling regime of a three dimensional 
non-degenerate attractive
Hubbard model was obtained in Ref.\cite{mic2}. Comparing our
results with theirs, one can immediately write down the $T_B$
for our case, and it is
\begin{equation}
k_{B}T_{B}=\frac{
2(n-2)\bar{q}(n)Zt^{2}_{P}}{2|U_A|ln(\frac{n}{4-n})}\,.
\end{equation}
The above formula is relevant for $|U_A|/(\sqrt{\bar{q}(n)}\,t_P)\geq 12$.
Here it must be pointed out that the finite temperature 
analysis in this section was done in real space and hence
one need not worry about the orthogonality of the Gutzwiller
projected excited state wave functions obtained by varying the occupation
numbers $n_{k\alpha\sigma}$, as was done in Ref.\cite{rice}.
Note that for $n=2$, we have a Mott insulator and $T_B$ is zero. 
While making $T_B$ estimates, we have to again
satisfy the constraints (i)-(iii)
given in the previous section. The small polaron formation
condition in the strong coupling regime is $g^{2}\omega>\sqrt{\bar{q}(n)}
D_{\circ}$ where $D_\circ$ is the original half-band-width equal
to $6t$ for a simple cubic lattice we have considered. The variation
of $T_B$ as a function of $\delta$ for various values of $\omega$
is shown in Fig. 11, and it is seen that decreasing $\omega$
increases the $T_B$'s and contracts the range of $\delta$ in which
Bose condensation is possible. The upper cut-off in Figs. 11-13 is
due to the violation of the small polaron condition. In Fig. 12,
we have displayed $T_B$ vs. $\delta$ for different values of $D_\circ$.
Decreasing $D_\circ$ decreases $T_B$ while expanding the $\delta$
range for Bose condensation. Finally, $T_B$ vs. $\delta$
for various values of $g$ is shown in
Fig. 13. Increase of $g$ is seen to sharply reduce $T_B$.
Overall, it is clear from the figures that $T_B$'s up to $100$ K can
be possible in the strong coupling regime. 
Furthermore, we note that slightly higher values of $T_B$ can be
obtained in the intermediate coupling regime as
compared to the strong coupling regime threshold ($|U|/t\approx 12$)
(see Fig. 2 in Ref.\cite{beck}).
\section{The Mott insulator with singlets}
\label{sect5}
We have noted earlier that the ground state of the Hamiltonian $H_P$ is
a Mott insulator for the case of two electrons per site. This Mott
insulating state obtains independent of the value of $U_A$ ($< 0$)
(see also the note in Ref.\cite{note}). 
The insulating state is obtained because of the strong
{\em inter}orbital repulsion $U_R$ between the JT polarons. The repulsion $U_R$
narrows the polaron band as the density increases  and eventually
drives it zero at $n=2$. Now, due to the
presence of finite {\em intra}orbital attraction between the polarons,
the insulating state
has a polaron singlet occupying one of the orbitals on  site. Since
the singlet can occupy either of the two orbitals available on  a site,
the insulting ground state is highly degenerate when $U_R$ is infinity.
When $U_R$ is large and finite, a pair on a site can gain energy by
virtual hopping to a nearest neighbor vacant orbital. 
This gain in energy, in second order perturbation theory, is 
$(-Zt^{2}_{P})/(2U_{R}+|U_A|)$ per polaron. Hence for finite and large $U_R$
the degeneracy of the insulating state reduces to two and the singlets
order in alternate orbitals on NN sites and thus we have an Orbital
Density Wave state.
We also note that one can map the paired polaron Mott insulator to an
Ising model since at each site the pair could be in either of the orbitals.
Assigning pseudo-spins $\uparrow$ to occupation of orbital 1 and 
$\downarrow$ to that of orbital 2, the Ising model is $H=\sum_{<i,j>}J
S_{i}.S_{j}$ with $J=t^{2}_{P}/(2U_{R}+|U_A|)$. The ground state corresponds
to the orbitally antiferromagnetic
state mentioned earlier.
When the electron density deviates
from $n=2$, the insulator undergoes an insulator to metal transition
and the superconducting instability of this metallic state was analyzed
in the previous sections.
\section{Conclusions}
\label{sect6}
In this paper we presented a theoretical study 
of superconductivity in a three dimensional
molecular conductor in which the molecular orbitals participating
in the conduction band formation
are doubly degenerate and EMVC for electrons in these orbitals interacting
with non-degenerate {\em intra}molecular vibration mode is large
enough to lead to the formation of Jahn-Teller (JT) small polarons.
We argued that the effective polaron-polaron interaction can be
attractive for material parameters realizable in molecular conductors,
and this interaction is the source of superconductivity in our model.
On analyzing superconducting instability in the weak and strong
coupling regimes of this attractive interaction,
we found that superconducting transition temperatures 
up to $100$ K are
achievable in molecular conductors within this mechanism.
We also find, for two particles per molecular site, that the
ground state is a novel Mott
insulating state in which a polaron singlet occupies one of
the doubly degenerate orbitals on each site. 
In the infinite {\em inter}orbital repulsion case, this Mott insulator
is highly degenerate since the singlet has the freedom to
occupy either of the orbitals on a site. On the other band, in 
the case of large but finite {\em inter}orbital repulsion case,
we found that the degeneracy of the ground state is reduced
to two with
the singlets orbitally antiferromagnetically ordered.
When the number of particles per site deviates from two,
the Mott insulator undergoes an insulator to metal transition.
\acknowledgements
One of the authors (S. Y.) would like to thank Sanjoy Datta for useful
discussions.
\appendix
\section{}
Here we show that, for strong coupling ({\em i.e}., $2G/\omega \gg 1$),
$E\otimes e$ JT
model is equivalent to the $E\otimes \beta$ 
model in Eq. (1). For $E\otimes e$, 
\begin{eqnarray}
\displaystyle
H&=&t\sum_{ij}\sum_{\alpha\sigma}
d^{\dag}_{i\alpha\sigma}d_{j\alpha\sigma}
+\frac{M\omega^2}{2}\sum_{i\alpha}Q^{2}_{i\alpha}
+\frac{M}{2}\sum_{i\alpha}\dot{Q}^{2}_{i\alpha}  \nonumber \\
&-&G\sqrt{2M\omega}\sum_{i\sigma}
\left(d^{\dag}_{i1\sigma}\,\,d^{\dag}_{i2\sigma}\right)
\left(
\begin{array}{rr}
Q_{1i} & Q_{2i} \\
Q_{2i} & -Q_{1i}
\end{array}\right)
\left(\begin{array}{c} d_{i1\sigma}\\
d_{i2\sigma}\end{array}\right)\nonumber \\
&+&U\sum_{j\alpha}n_{j\alpha\uparrow}n_{j\alpha\downarrow}
+U\sum_{j\sigma\sigma^{\prime}}
n_{j1\sigma}n_{j2\sigma^{\prime}}\,.
\end{eqnarray}
Now we do a rotation
transformation on $H$. For this purpose, let
\begin{equation}
-\left(
\begin{array}{rr}
Q_{1i} &  Q_{2i} \\
Q_{2i} & -Q_{1i}
\end{array}\right)=Q_{i}
\left(
\begin{array}{rr}
-cos(\theta_{i}) & sin(\theta_{i}) \\
 sin(\theta_{i}) & cos(\theta_{i})
\end{array}\right),
\end{equation}
where $Q_{i}=\sqrt{Q^{2}_{1i}+Q^{2}_{2i}}$ and
\begin{equation}
\left(\begin{array}{c} c_{i1\sigma}\\
\\
c_{i2\sigma}\end{array}\right)=
\left(
\begin{array}{rr}
sin(\frac{\theta_{i}}{2}) & cos(\frac{\theta_{i}}{2}) \\
\\
cos(\frac{\theta_{i}}{2}) & -sin(\frac{\theta_{i}}{2})
\end{array}\right)
\left(\begin{array}{c} d_{i1\sigma}\\
\\
d_{i2\sigma}\end{array}\right)\,.
\end{equation}
Using the above, $H$ is transformed to:
\begin{eqnarray}
\displaystyle
H&=&t\sum_{ij\sigma}
\left(c^{\dag}_{i1\sigma}\,\,c^{\dag}_{i2\sigma}\right)
\left(
\begin{array}{rr}
1 & 0 \\
0 & 1
\end{array}\right)
\left(\begin{array}{c} c_{j1\sigma}\\
c_{j2\sigma}\end{array}\right)\nonumber \\
&+&G\sqrt{2M\omega}\sum_{i\sigma}
\left(c^{\dag}_{i1\sigma}\,\,c^{\dag}_{i2\sigma}\right)
\left(\begin{array}{rr}
Q_{i} & 0 \\
0 & -Q_{i}
\end{array}\right)
\left(\begin{array}{c} c_{i1\sigma}\\
c_{i2\sigma}\end{array}\right)\nonumber \\
&+&\frac{M\omega^2}{2}\sum_{i}Q^{2}_{i}
+\frac{1}{2M}
\sum_{i}\left(\frac{\partial^{2}}{\partial Q^{2}_{i}}+
\frac{1}{Q_{i}}\frac{\partial}{\partial Q_{i}}+
\frac{1}{Q^{2}_{i}}\frac{\partial^{2}}{\partial\theta^{2}_{i}}
\right) \nonumber \\
&+& U\sum_{j\alpha}n_{j\alpha\uparrow}n_{j\alpha\downarrow}
+U\sum_{j\sigma\sigma^{\prime}}
n_{j1\sigma}n_{j2\sigma^{\prime}}\,.
\end{eqnarray}
As argued in Ref.\cite{sturge}, for $2G/\omega\gg 1$,
$(1/Q_{i})({\partial}/{\partial Q_{i}})$ and
$(1/Q^{2}_{i})(\partial^{2}/\partial\theta^{2}_{i})$
are unimportant, and we are left with  an effective
single mode equation. Then using $\sqrt{2M\omega}Q_{i}=b_{i}+b^{\dag}_{i}$,
one gets the Eq. (1) given in the text.
Here it should be pointed out that the single normal mode approximation
for $E\otimes e$ works only for the low lying excitations of the mode.
For the transition temperatures obtained and the frequencies 
considered $\omega/k_{B}T_{c} >> 1$, and hence phonon 
averaging after the Lang-Firsov transformation yields
for the single mode approximated $E\otimes e$ system the
same result given by Eq. (15) in the text.

\begin{figure}
\resizebox*{3.1in}{2.5in}{\rotatebox{0}{\includegraphics{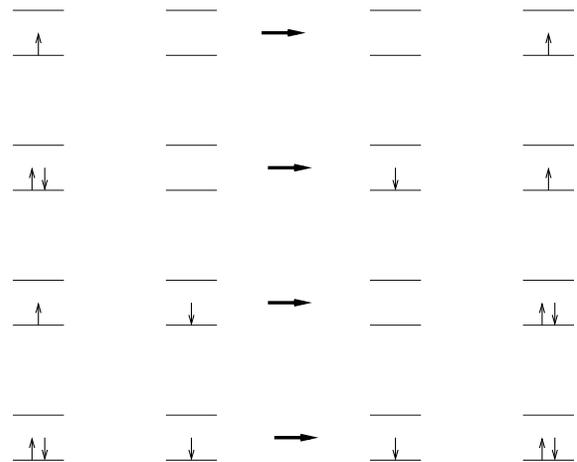}}}
\vspace*{0.5cm}
\caption[]{The inter-site hopping processes considered (see text).}
\label{scaling}
\end{figure}
\begin{figure}
\resizebox*{3.1in}{2.5in}{\rotatebox{270}{\includegraphics{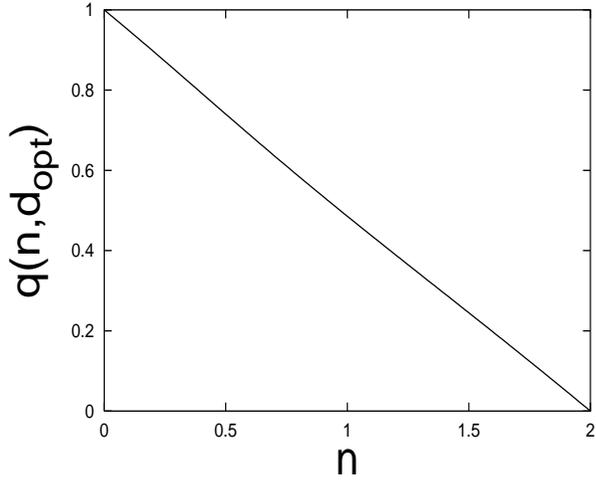}}}
\vspace*{0.5cm}
\caption[]{The variation of the band-width renormalization
factor as a function of particle density.}
\label{scaling}
\end{figure}
\begin{figure}
\resizebox*{3.1in}{2.5in}{\rotatebox{270}{\includegraphics{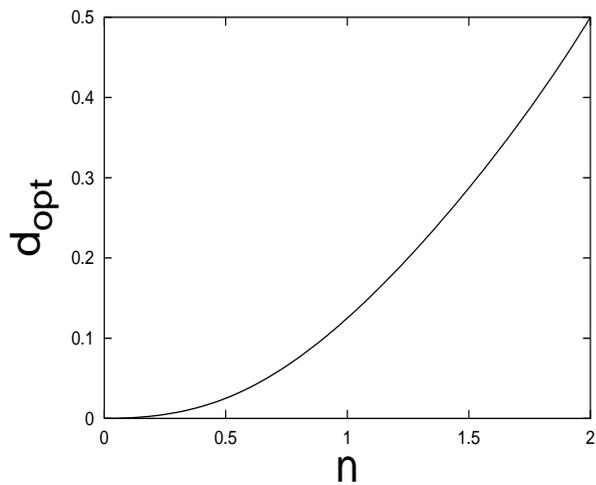}}}
\vspace*{0.5cm}
\caption[]{The variation of the {\em intra}orbital double occupancy
as a function of particle density.}
\label{scaling}
\end{figure}
\begin{figure}
\resizebox*{3.1in}{2.5in}{\rotatebox{270}{\includegraphics{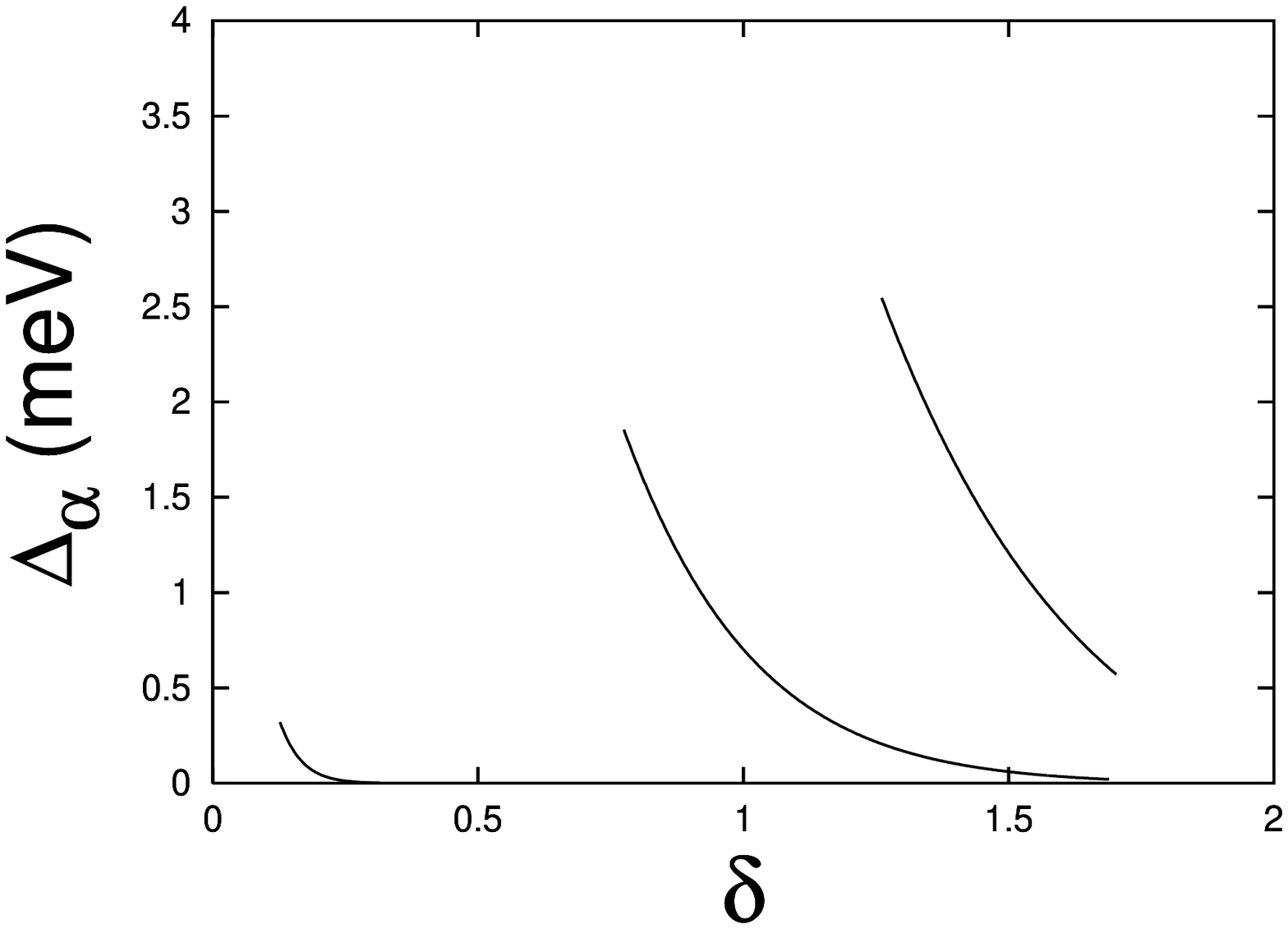}}}
\vspace*{0.5cm}
\caption[]{$\Delta_{\alpha}$ vs. $\delta$ for various
values of $\omega$ (in eV): 0.348 (left),
0.352 (middle), 0.355 (right). Values of the other
parameters are: $U=1\,eV$, $D_{\circ}=0.6\,eV$,
and $g=1.2$.}
\label{scaling}
\end{figure}
\begin{figure}
\resizebox*{3.1in}{2.5in}{\rotatebox{270}{\includegraphics{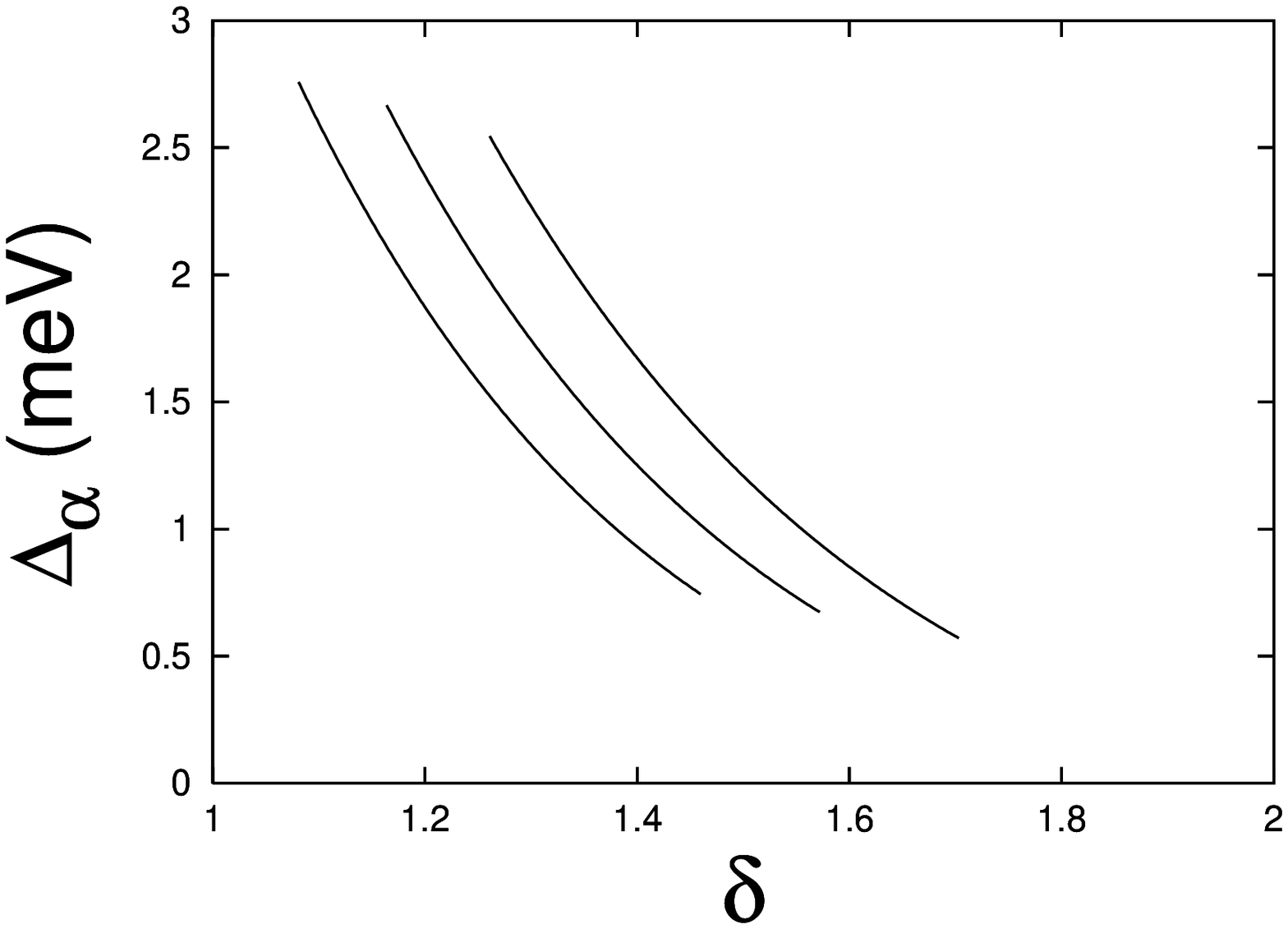}}}
\vspace*{0.5cm}
\caption[]{$\Delta_{\alpha}$ vs. $\delta$ for various
values of $D_{\circ}$ (in eV): 0.70 (left),
0.65 (middle), 0.60 (right). Values of the other
parameters are: $U=1\,eV$, $\omega=0.355\,eV$,
and $g=1.2$.}
\label{scaling}
\end{figure}
\begin{figure}
\resizebox*{3.1in}{2.5in}{\rotatebox{270}{\includegraphics{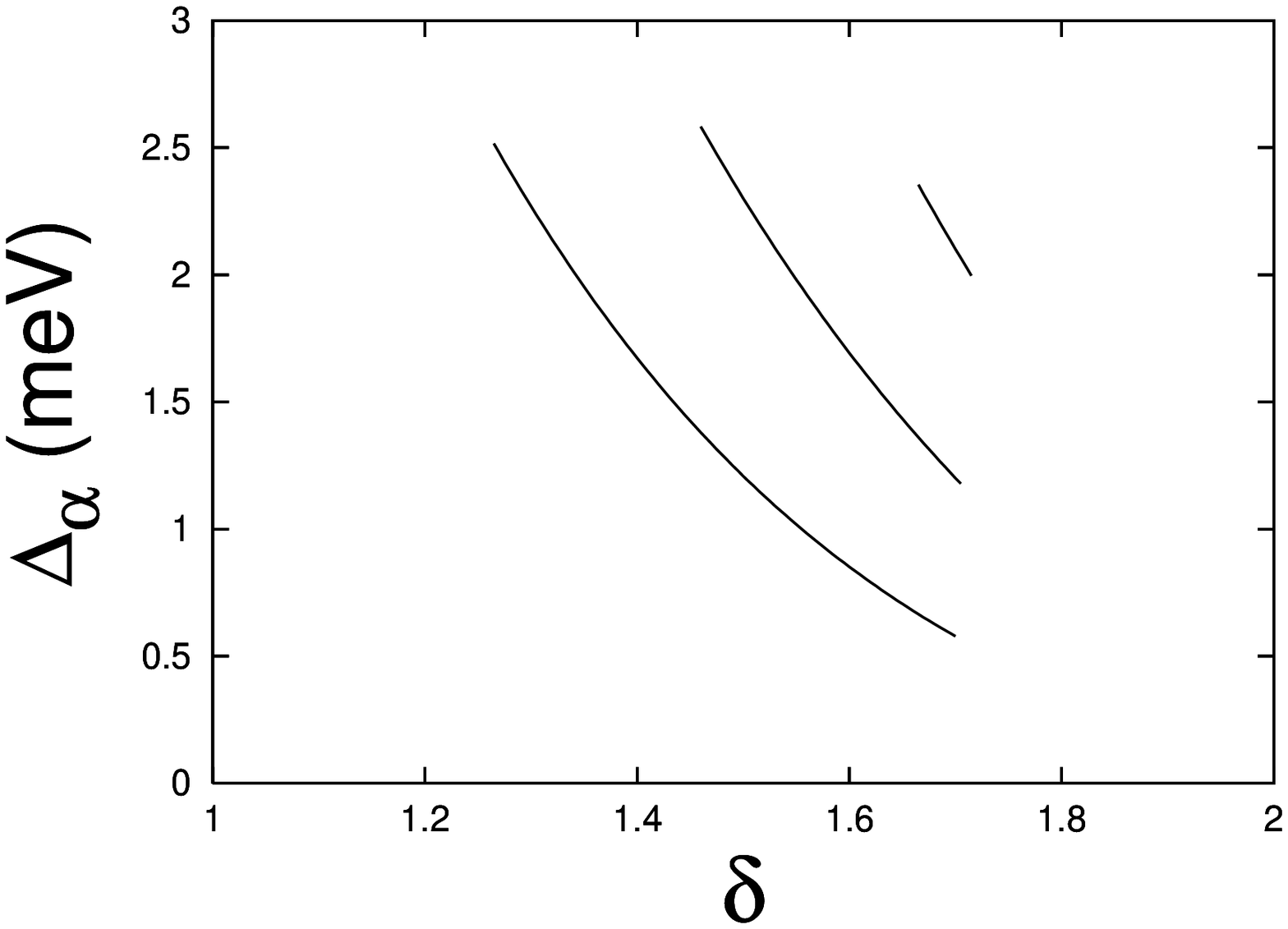}}}
\vspace*{0.5cm}
\caption[]{$\Delta_{\alpha}$ vs. $\delta$ for various
values of $g$: 1.200 (left),
1.202 (middle), 1.204 (right). Values of the other
parameters are (in eV): $U=1$, $\omega=0.355$,
and $D_{\circ}=0.6$.}
\label{scaling}
\end{figure}
\begin{figure}
\resizebox*{3.1in}{2.5in}{\rotatebox{270}{\includegraphics{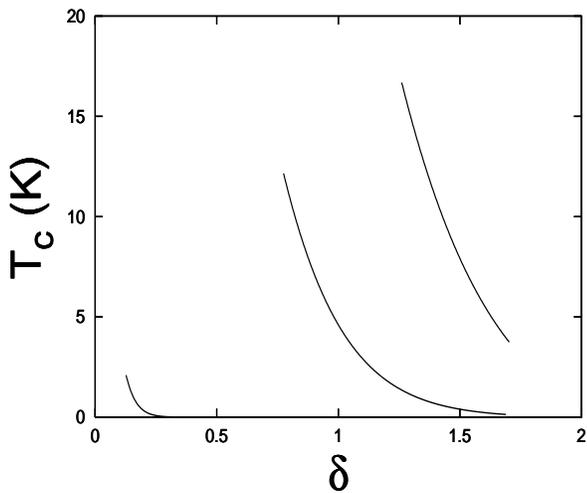}}}
\vspace*{0.5cm}
\caption[]{T$_{c}$ vs. $\delta$ for various
values of $\omega$ (in eV): 0.348 (left), 
0.352 (middle), 0.355 (right). Values of the other 
parameters are: $U=1\,eV$, $D_{\circ}=0.6\,eV$,
and $g=1.2$.}
\label{scaling}
\end{figure}
\begin{figure}
\resizebox*{3.1in}{2.5in}{\rotatebox{270}{\includegraphics{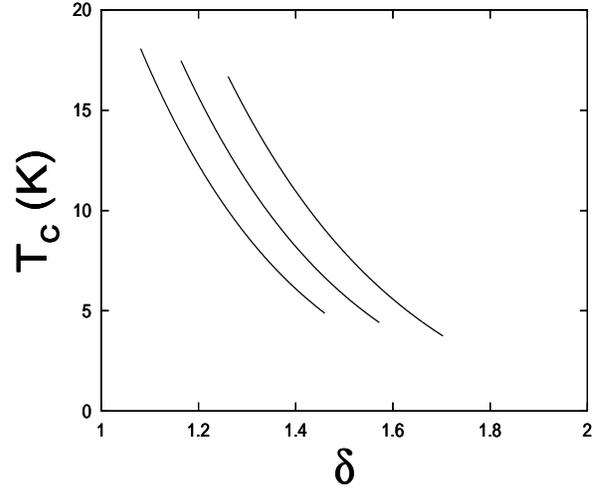}}}
\vspace*{0.5cm}
\caption[]{T$_{c}$ vs. $\delta$ for various
values of $D_{\circ}$ (in eV): 0.70 (left), 
0.65 (middle), 0.60 (right). Values of the other
parameters are: $U=1\,eV$, $\omega=0.355\,eV$,
and $g=1.2$.}
\label{scaling}
\end{figure}
\begin{figure}
\resizebox*{3.1in}{2.5in}{\rotatebox{270}{\includegraphics{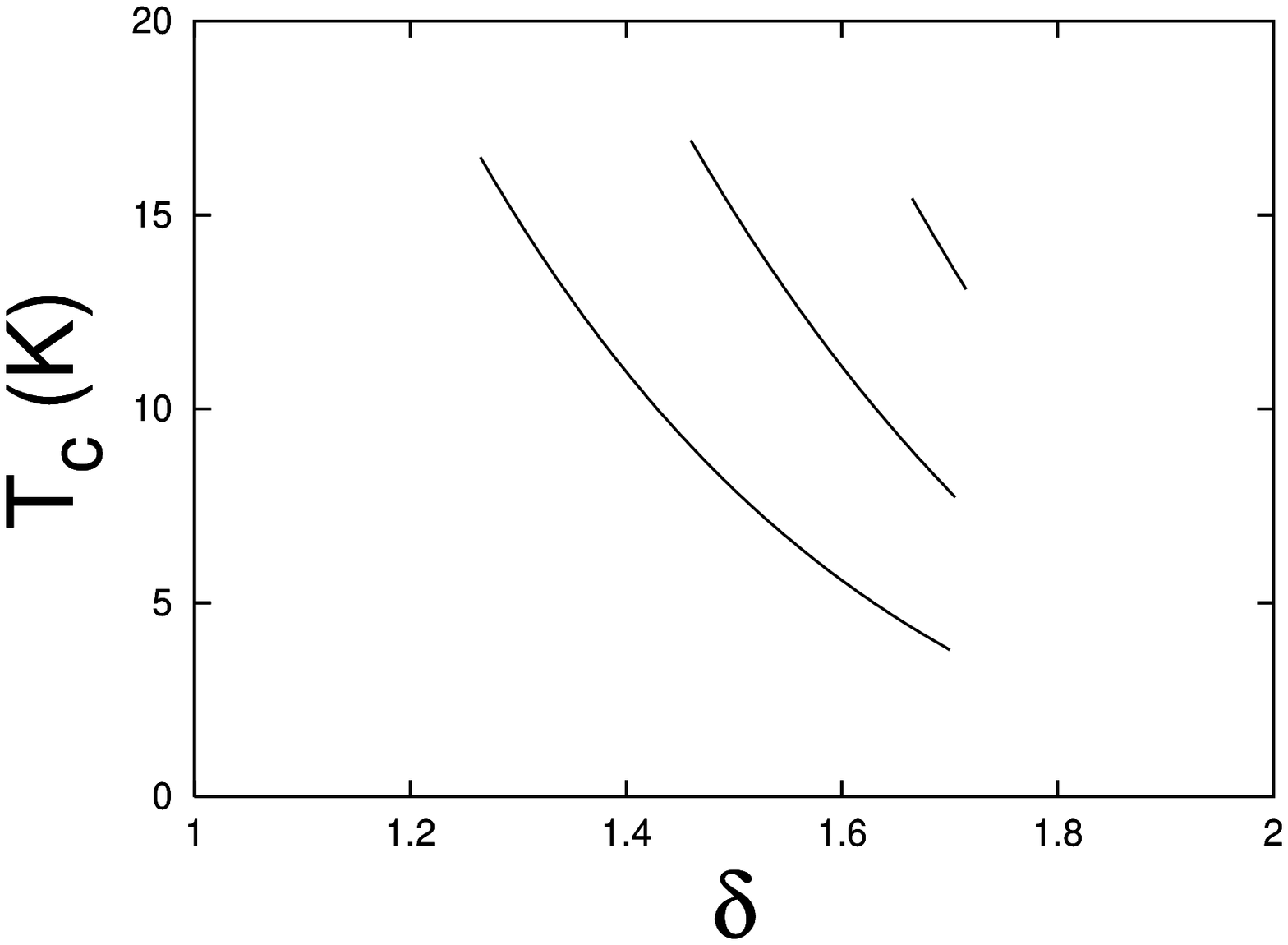}}}
\vspace*{0.5cm}
\caption[]{T$_{c}$ vs. $\delta$ for various
values of $g$: 1.200 (left),       
1.202 (middle), 1.204 (right). Values of the other
parameters are (in eV): $U=1$, $\omega=0.355$,
and $D_{\circ}=0.6$.}
\label{scaling}
\end{figure}
\begin{figure}
\resizebox*{3.1in}{2.0in}{\rotatebox{0}{\includegraphics{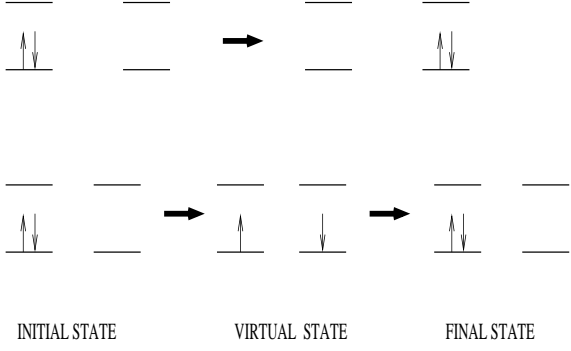}}}
\vspace*{0.5cm}
\caption[]{The inter-site hopping processes (see text) in
the Bose regime.}
\label{scaling}
\end{figure}
\begin{figure}
\resizebox*{3.1in}{2.5in}{\rotatebox{270}{\includegraphics{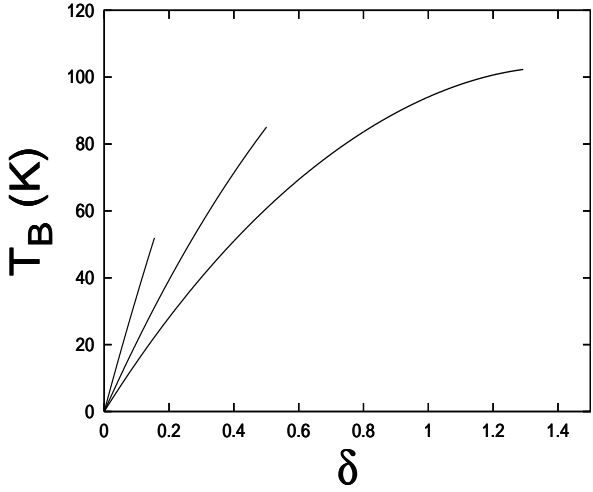}}}
\vspace*{0.5cm}
\caption[]{T$_{B}$ vs. $\delta$ for various
values of $\omega$ (in eV): 0.350 (left),
0.375 (middle), 0.400 (right). Values of the other
parameters are: $U=0.9\,eV$, $D_{\circ}=0.6\,eV$,
and $g=1.2$.}
\label{scaling}
\end{figure}
\begin{figure}
\resizebox*{3.1in}{2.5in}{\rotatebox{270}{\includegraphics{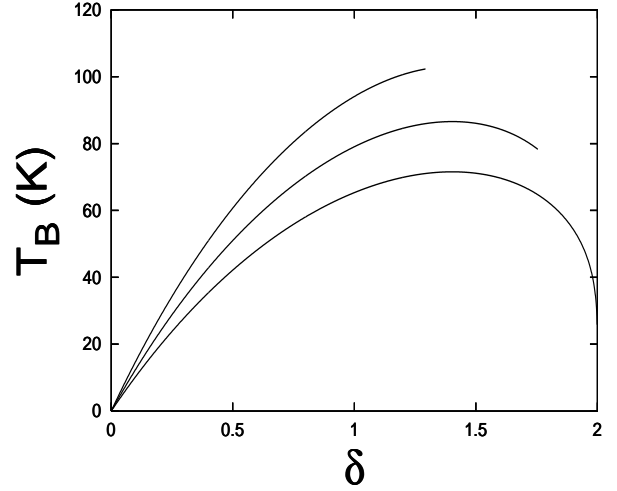}}}
\vspace*{0.5cm}
\caption[]{T$_{B}$ vs. $\delta$ for various
values of $D_{\circ}$ (in eV): 0.60 (top),
0.55 (middle), 0.50 (bottom). Values of the other
parameters are: $U=0.9\,eV$, $\omega=0.4\,eV$,
and $g=1.2$.}
\label{scaling}
\end{figure}
\begin{figure}
\resizebox*{3.1in}{2.5in}{\rotatebox{270}{\includegraphics{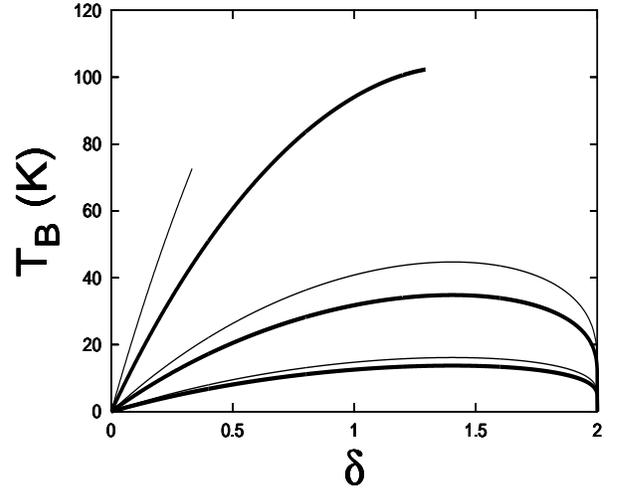}}}
\vspace*{0.5cm}
\caption[]{T$_{B}$ vs. $\delta$ for various
values of $g$: 1.20 (top),
1.3 (middle), 1.4 (bottom). Values of the other
parameters are (in eV): $U=0.9$ (thick lines) and
$U=1.0$ (thin lines), $\omega=0.4$,
and $D_{\circ}=0.6$.}
\label{scaling}
\end{figure}
\end{multicols}

\begin{references}
\bibitem{note}
For $n=2$, a Mott insulating state is also possible
for $0\leq U_{A}\ll U_R$.
\end{references}
\end{document}